\documentclass[aps,pra,twocolumn,amssymb,amsfonts,floatfix,showpacs]{revtex4}

\usepackage{graphicx}% Include figure files
\usepackage{dcolumn}% Align table columns on decimal point
\usepackage{bm}% bold math
\usepackage{mathbbol}
\usepackage[dvips]{color}
\usepackage{amsmath,amssymb,amsthm}
\usepackage{color}

\begin{document}

\title{Domain wall dynamics in integrable and chaotic spin-1/2 chains}

\author{Lea F. Santos}
\email{lsantos2@yu.edu}
\affiliation{Department of Physics, Yeshiva University, 
245 Lexington Ave, New York, NY 10016, USA}
\author{Aditi Mitra}
\email{aditi.mitra@nyu.edu}
\affiliation{Department of Physics, New York University, 
4 Washington Place, New York, NY 10003, USA}

\begin{abstract}
We study the time evolution of 
correlation functions, spin current, and local magnetization in an isolated spin-1/2 chain 
initially prepared in a sharp domain wall state. The results are compared with the 
level of spatial delocalization of the eigenstates of the system which is measured 
using the inverse participation ratio. 
Both integrable and non-integrable regimes are 
considered. Non-integrability is introduced to the integrable Hamiltonian with nearest neighbor couplings
by adding a single site impurity field or by adding next-nearest-neighbor couplings.
A monotonic correspondence between the enhancement of the level of delocalization,
spin current and magnetization dynamics   
occurs in the integrable domain. This correspondence is however
lost for chaotic models with weak Ising interactions. 
\end{abstract}

\date{\today}
\maketitle

\section{Introduction}

The dynamical behavior of quantum many-body systems out of equilibrium 
offers several open questions. 
In terms of transport properties, for example, the conditions for ballistic or
diffusive transport have received a good deal of attention throughout the years~\cite{Zotos1997,Heidrich2003,Alvarez2002,Heidrich2004,Zotos2005,Jung2006,Jung2007,Mukerjee2008,Langer2009,Barisic2009,Sirker2011}. The interest
in the subject has been further reinforced by experiments with magnetic compounds, 
where ballistic transport of heat has been 
reported~\cite{Sologubenko,Hess}.
Another stimulating scenario appears in the context of experiments with optical lattices~\cite{Greiner2008,Bloch2008}, where the dynamics
of strongly correlated quantum systems can be investigated for long times.
Here, the possibility of studying the problem of thermalization in isolated quantum
systems~\cite{kinoshita06} has in part motivated the renewed interest in this topic
of research~\cite{rigol08STATc,PolkovnikovARXIV}.

In the present work we are interested in the quench dynamics of 
one-dimensional systems of interacting spins-1/2. We focus on 
a particular initial state corresponding to a
sharp domain wall where the spins point up in the first half 
of the chain and down in the other half.
The experimental realization of such inhomogeneous states requires
a magnetic field gradient, which has been realized 
in an optical lattice loaded with
atoms in two hyperfine states~\cite{Weld2009}.
The domain wall state was considered before in numerical studies
of magnetization transport in antiferromagnetic
spin-1/2 XXZ chains~\cite{Gobert2005}, where using the adaptive time-dependent density-matrix 
renormalization group method, a sharp suppression of transport was verified for large interactions. 
It also appeared in studies of the dynamics of a spinon in an external magnetic 
field~\cite{Cai2011} and was used in numerical comparisons of transport behavior in integrable
and nonintegrable systems~\cite{Steinigeweg2006,Santos2008PRE,SantosLocMag}.
In the context of thermalization, the
dynamics of a sharp domain wall in the gapped phase of the XXZ chain 
was studied using Algebraic Bethe Ansatz in~\cite{Mossel2010}.

The evolution of a domain wall state was investigated analytically 
in~\cite{Antal1999} and~\cite{Lancaster2010} for the XX model and  
numerically employing the truncated Wigner approximation 
for the quantum sine-Gordon model~\cite{Lancaster2010a}. 
The domain wall in the XX model and the Luttinger liquid was found to 
spread out ballistically into a current carrying steady state~\cite{Lancaster2010,Lancaster2010a}. 
In a spin-chain, current    
flow implies a rotation in space of the transverse component of the spins. 
As a result, the transverse spin correlation function in these simple models
were found to reach 
a nonequilibrium steady state which showed spatial oscillations at a wavelength inversely 
related to the magnitude of the current~\cite{Lancaster2010}.
As back-scattering interactions were added, the current was found to decrease,
and consequently, the wavelength of the oscillations in the transverse spin
correlation function was found to increase~\cite{Lancaster2010a}.
These results are further explored here taking into account both
integrable and nonintegrable systems. The approach is to study the time-evolution
numerically employing EXPOKIT~\cite{Expokit,Sidje1998}. 

EXPOKIT is a software package based on Krylov subspace projection methods.
Instead of diagonalizing the complete system Hamiltonian $H$, we use this package
to directly compute the action of the matrix exponential $e^{-iHt}$ on a vector
of interest. We consider spin-chains with short range
couplings and $20$ sites leading to 
sparse matrix Hamiltonians of dimension $184\,756$. 

Based on transverse spin correlation function, spin current, and local magnetization, 
we perform a numerical comparison between 
integrable spin-1/2 chains with nearest-neighbor couplings only and their nonintegrable counterparts. The 
latter are realized 
by adding a single-site impurity field or by adding next-nearest-neighbor couplings.  
The wavelength of the oscillations of the transverse spin correlation function
increases with the Ising interaction for the three models.
More unexpected are the results for current and magnetization for the chaotic models.
For the integrable model, the behavior is predictable. As the Ising interaction
increases, the eigenstates of the system become spatially more localized, 
which reflects in the  motion of the excitations: 
the spin current decreases and the decay of the absolute
local magnetization slows down. For 
large interactions, the same tendency is verified for the chaotic models, 
whereas surprises appear as the strength $\Delta $ 
of the Ising interaction changes from zero to 
small values. For the model with impurity, the addition of a
small interaction leads to the onset of chaos, which is followed by
a significant increase of the level of spatial delocalization of the eigenstates.
Contrary to that, the magnitude of the current and the decay rate of the 
magnetization decrease.
In the case of next-nearest-neighbor couplings, the system is chaotic
even in the absence of Ising interaction
(provided that the ratio between next-nearest- and nearest-neighbor couplings is 
sufficiently large). Adding $\Delta $ contracts the eigenstates, 
whereas, quite counterintuitively, 
the spin current increases and the decay of the absolute local
magnetization is enhanced.

For the model with impurity, the opposite behavior between 
delocalization and dynamical spreading of excitations may 
be an indication of the transition from ballistic
to diffusive transport. We do not have a complete 
explanation for the behavior of the chaotic
system with next-nearest-neighbor couplings, but 
hope this work will motivate further investigation.

The paper is organized as follows. In Sec.~\ref{model}, the models and
the physical quantities studied are defined and a description 
of the numerical method is provided. In Sec.~\ref{results},
the results for the correlation function, magnetization and current are
analyzed, and in Sec.~\ref{ipr}, these results are
compared with the level of spatial delocalization of the eigenstates of the system
which is measured in terms of the inverse participation ratio.
Concluding remarks are presented in Sec.~\ref{concl}.

\section{Models and numerical method} \label{model}

We study a one-dimensional spin-1/2 system with open boundary conditions described by the Hamiltonian

\begin{equation}
H = H_z + H_{\text{NN}}+ \alpha H_{\text{NNN}} ,
\label{ham} 
\end{equation}
where

\begin{align}
&H_z =  \epsilon_{L/2+1} S_{L/2+1}^z , 
\label{Hz} \\
&H_{\text{NN}} = \sum_{n=1}^{L-1} J\left[ \left( 
S_n^x S_{n+1}^x + S_n^y S_{n+1}^y \right) +
\Delta S_n^z S_{n+1}^z \right] ,
\label{HNN} \\
&H_{\text{NNN}} = \sum_{n=1}^{L-2} J\left[ \left( S_n^x S_{n+2}^x + S_n^y S_{n+2}^y
\right) + \Delta S_n^z S_{n+2}^z  \right] \:.
\label{HNNN}
\end{align}
Above,  $\hbar$ is set equal to 1,
$L$ is the number of sites, and $S^{x,y,z}_n = \sigma^{x,y,z}_n/2$ 
are the spin operators at site $n$,  $\sigma^{x,y,z}_n$ being the Pauli matrices. 

(i) The system is clean when all Zeeman splittings, 
determined by a static magnetic field in the $z$ direction,
are equal and it has on-site disorder when the spins have different energy splittings. 
We will consider the 
case of a single-site impurity field where only the Zeeman splitting $\epsilon_{L/2+1}$
at site $L/2+1$ is different from zero.

(ii) The XXZ Hamiltonian $H_{\text{NN}}$ describes the nearest-neighbor 
(NN) exchange; $S_n^z S_{n+1}^z$ is the Ising interaction and 
$S_n^x S_{n+1}^x + S_n^y S_{n+1}^y$ is the flip-flop term which 
tends to orient the spins in the $XY$ plane.
The coupling strength $J$ and 
the anisotropy $\Delta$ are assumed positive, thus favoring antiferromagnetic order. 
By varying $\Delta$ in $H=H_{NN}$, the system moves from 
the gapless $XY$ phase ($\Delta < 1$)
to the gapped Ising phase ($\Delta > 1$), which, at $T=0$, corresponds to 
the metal-insulator Mott transition \cite{Cloizeaux1966}. 
In the particular case of $\Delta=0$, we deal with 
the XX model; whereas $\Delta=1$ corresponds to the 
isotropic Heisenberg Hamiltonian~\cite{SutherlandBook}.

(iii) The parameter
$\alpha$ refers to the ratio between the next-nearest-neighbor (NNN) 
exchange, as determined by $H_{\text{NNN}}$, and the NN couplings.
The inclusion of NNN antiferromagnetic exchange 
frustrates the chain, since NN exchange favors ferromagnetic alignment between the second neighbors.

Depending on the parameters of Hamiltonian (\ref{ham}), different symmetries
are found. The total spin in the $z$ direction, $S^z=\sum_{n=1}^{L} S_n^z$, 
is conserved for all parameters.
When $\epsilon_{L/2+1}=0$, 
other symmetries include~\cite{Brown2008}:  invariance under lattice reflection, 
which leads to parity conservation; 
invariance under a $\pi$-rotation around the $x$-axis in the case of $S^z=0$; and
conservation of total spin, $S^2=(\sum_{n=1}^L \vec{S}_n)^2$ when $\Delta=1$. 

\subsection{Integrable and chaotic regimes}

Adjusting the parameters of the Hamiltonian, the system may undergo a crossover to the 
chaotic regime. We consider here the following scenarios.

$\bullet$ {\em Integrable model with $\epsilon_{L/2+1}=0.0, \alpha=0.0$; 
$H=H_{\text{NN}}$}. 
A clean one-dimensional spin-1/2 model 
with NN couplings only is 
integrable. When $\Delta=0$, it can be mapped onto a system of 
noninteracting spinless fermions~\cite{Jordan1928}. 
When $\Delta \neq 0$, its ground state properties can be studied using
Bethe Ansatz~\cite{Bethe}.

$\bullet$ {\em Chaotic model with $\epsilon_{L/2+1}\neq 0, \alpha=0.0$;
$H=H_z+H_{\text{NN}}$}. 
The addition of on-site disorder~\cite{Avishai2002} to the XXZ model, 
even if it corresponds to a single impurity in the middle of the chain~\cite{Santos2004},
leads to the onset of quantum chaos provided $\epsilon_{L/2+1} \lesssim J$ (if 
the impurity becomes too large it breaks the chain in two pieces)
and $\Delta \leq 1$, but not zero. The presence of interaction is 
essential; it is the interplay between interaction and disorder that leads
to chaos. Here, we will consider an impurity of strength $\epsilon_{L/2+1}=0.5J$.

$\bullet$ {\em Chaotic model with $\epsilon_{L/2+1}=0.0, \alpha\neq 0$;
$H=H_{\text{NN}}+\alpha H_{\text{NNN}}$}. 
The inclusion of
NNN couplings may also drive the system to the chaotic domain
for sufficiently large values of $\alpha$ \cite{Hsu1993,Kudo2005,notePRE}. 
We will consider $\alpha=0.4$, which guarantees chaoticity when $\Delta \leq 1$.
It is interesting that the system is chaotic even when $\Delta=0$,
which we verified by computing the level spacing distribution~\cite{Guhr1998}. 
When $\alpha \neq 0$ and $\Delta=0$, the Jordan-Wigner transformation 
does not map the spin-1/2 chain onto a system of noninteracting spinless fermions, but instead
correlated hopping terms appear, which is the cause for the onset of chaos.
The level spacing distribution is the most common way to verify whether
a system becomes chaotic. It requires the complete diagonalization 
of the Hamiltonian and, very importantly, the separation of the eigenvalues according to
symmetry sectors. If eigenvalues from different subspaces are mixed, we may obtain a 
Poisson distribution (indicating integrability) even if the system is chaotic. 

\subsection{Quantities of interest}

In Sec.~\ref{results}, we will study the dynamical properties of three quantities 
of interest, which are described below.

(i) The transverse spin correlation function corresponds to

\begin{equation}
C^{xx}(n,n+m) = \langle S^x_n S^x_{n+m} \rangle .
\label{cxx} 
\end{equation}

(ii) The local spin current, $I_{s,n}$, is associated with 
the conservation of total spin in the $z$ direction, $S^z=\sum_{n=1}^{L} S_n^z$
and obeys the continuity equation,

\[
\frac{\partial S^z_n}{\partial t} + div(I_{s,n})=0.
\]
In the bulk \cite{closed},
\[
-i [H, S_n^z] = div(I_{s,n})= (I_{s,n} - I_{s,n-1}),
\]
but in the extremes, since the chain is open, we have
\[
-i [H, S_1^z] =div(I_{s,1})=I_{s,1} 
\]
and
\[
-i [H, S_L^z] =div(I_{s,L})= -I_{s,L-1}.
\]

From the equations above, we find that the local spin current for
a system with NN couplings only, with or without impurity,
is given by

\begin{equation}
I_{s,n}^{(1)} = J (S_n^x S_{n+1}^y - S_n^y S_{n+1}^x),
\label{I1}
\end{equation}
where $1\leq n \leq L-1$.

In the case of the model with NN and NNN couplings, the local spin current becomes

\begin{eqnarray}
I_{s,n}^{(2)} &=& J(S_n^x S_{n+1}^y - S_n^y S_{n+1}^x)  
\label{I2} \\
&+&\alpha J (S_n^x S_{n+2}^y - S_n^y S_{n+2}^x
+S_{n-1}^x S_{n+1}^y - S_{n-1}^y S_{n+1}^x), \nonumber
\end{eqnarray}
for $2\leq n \leq L-2$, and at the borders,
\begin{align}
&I_{s,1}^{(2)}=J(S_1^x S_{2}^y - S_1^y S_{2}^x) + 
\alpha J (S_1^x S_{3}^y - S_1^y S_{3}^x), \nonumber \\
&I_{s,L-1}^{(2)}=J(S_{L-1}^x S_{L}^y - S_{L-1}^y S_{L}^x) 
+\alpha J (S_{L-2}^x S_{L}^y - S_{L-2}^y S_{L}^x). \nonumber
\end{align}

Equations.~(\ref{cxx}) and~(\ref{I1}) imply that for the
XXZ chain a nonzero current gives rise to a static spin configuration where 
the spins rotate in the
$XY$ plane from one site to the next at a rate which increases with the magnitude of the current. However, 
as is apparent in Eq.~(\ref{I2}), this
simple relation between the NN spin correlations and the current is
lost for NNN couplings.  

(iii) The local magnetization of each site is,
\begin{equation}
M_{n} = \langle \Psi(t)| S_n^z|\Psi(t) \rangle,
\label{Mn}
\end{equation}
while the local magnetization of the first half of the chain is given by,
\begin{equation}
M_{(1,L/2)} = \langle \Psi(t)|\sum_{n=1}^{L/2} S_n^z|\Psi(t) \rangle.
\label{M1}
\end{equation}

The results for the dynamical behavior of the above quantities will then 
be compared with the level 
of delocalization of the eigenstates in Sec.~\ref{ipr}.

\subsection{Numerical method}

We use EXPOKIT~\cite{Expokit,Sidje1998} to study numerically the 
time evolution of the system. This is a software package that 
can deal with large sparse matrices, as the ones we have in this work. It avoids
the complete diagonalization of the Hamiltonian and instead provides routines to compute
the matrix exponential $e^{-iHt}$. 
The algorithm computes the matrix exponential times a vector $|\Psi\rangle $. 
The main idea is to approximate $e^{-iHt} |\Psi\rangle$ by 
an element of the Krylov subspace of dimension $k$, which is small compared to the dimension
of the total Hamiltonian. The Krylov technique is based on the 
Arnoldi iteration when dealing with a general matrix and on the Lanczos iteration 
if the matrix is symmetric or Hermitian.
The product $e^{-iHt} |\Psi\rangle$  is 
approximately evaluated within the low-dimensional Krylov subspace. The problem is then
reduced to the calculation of an exponential of a dense $k\times k$ matrix,
which is done by using the rational Pad\'e approximations.

Here, we consider $k=30$ and time steps of $0.01/J$. The software proved to be very fast 
and accurate. For a chain with 
$L=16$, for example, we computed the deviation $|M_{exact}-M_{expokit}|$ 
between the result  
for site magnetization 
obtained from exact diagonalization ($M_{exact}$), and that obtained with EXPOKIT ($M_{expokit}$),
for all sites of the chain.
This was done for all 3001 instants of time between 0 and $30/J$.
For a system with 8 up-spins, $\Delta=0.5$, and for both NN and NNN couplings
($\alpha=0.4$),
the deviation was always smaller than $10^{-8}$. Using complete exact diagonalization, it 
took almost 23 hours to get the results, whereas with EXPOKIT it took less than 8 minutes
on the same machine.

\section{Numerical results: comparison between integrable and nonintegrable models}
\label{results}

We consider $L$ even and choose as an initial state a
sharp domain wall where the spins point up in the first half of the chain and down in the 
other half, $| \Psi(0) \rangle=| \uparrow \uparrow \uparrow \ldots \uparrow \downarrow \ldots 
\downarrow \downarrow \downarrow \rangle$.   This state belongs to the 
largest subspace, $S^z=0$, which has dimension 
$\mbox{dim}={L \choose L/2}=L!/[(L/2)!]^2$. 

Experimentally such a domain wall state may be created by the application of a
strong spatially varying magnetic field. 
The time-evolution is initiated when this strong field is quenched 
and the couplings become suddenly effective. The system
dynamics is then dictated by Hamiltonian~(\ref{ham}) with $\epsilon_n=0$
for all sites or, in the case of an impurity, with a small field remaining
only on site $L/2+1$.

The decision to place the impurity at site $L/2+1$ and not at $L/2$ has a reason.
Placing it at $L/2$ implies the same results in terms of chaos indicators, but 
the transport of the excitations is impeded even before $\Delta$ surpasses 1.
Due to the Ising interaction and border effects,
the domain wall has a very large energy of $(L-3)\Delta/4$. As $\Delta $ increases, 
and especially when it becomes larger than 1, it becomes difficult to find 
states that are resonant with $|\Psi(0)\rangle$ and the domain wall hardly moves. This
localization occurs for even smaller values of $\Delta$ if the impurity
is at $L/2$, because it adds to the already large energy of the domain wall 
state an additional positive contribution from the impurity field~\cite{Santos2008PRE}. 

For the XX model, the width of the domain wall increases linearly in 
time~\cite{Lancaster2010}. As a result, the rightmost excitation
(the up-spin in the middle of the domain wall) reaches the 
border of the chain at time $t\sim L/(2J)$. Thus, 
to avoid border effects, we will focus on $t\leq L/(2J)$ for models
with NN couplings only,
even in the presence of interaction and disorder. 
In the case of systems with both NN and NNN couplings, the excitations
move faster (as can be seen in Fig.~\ref{fig3:contour}), 
so we will center our attention on times not much larger than $L/(4J)$.

\subsection{Transverse spin correlation function} 

We study the transverse spin correlation function,
Eq.~(\ref{cxx}), between site $L/2$ and 
sites $L/2+m$ after a time long enough for the domain wall to have broadened
and short enough to prevent effects from the borders.

It was analytically shown 
for the XX model 
that the correlation function at $t\rightarrow \infty$ 
reaches a nonequilibrium steady state
given by~\cite{Lancaster2010},

\begin{equation}
C^{xx}(n,n+m) \stackrel{t\rightarrow \infty}{\rightarrow}
C_{eq}^{xx}(m) \cos \left( \frac{2 \pi m}{\lambda} \right),
\label{cxx_XX}
\end{equation}
where the ground state correlation function is,
\begin{equation}
C_{eq}^{xx}(m) \approx \frac{1}{\sqrt{8 m}} \kappa^2, 
\label{equilibrium}
\end{equation}
with $\log \kappa = \frac{1}{4} \int_0^{\infty} \frac{dt}{t} \left(
e^{-4 t} - \frac{1}{\cosh^2 t}\right)$.
Thus the ground state correlation function is 
modulated by oscillations at a wavelength $\lambda$ where 
$\lambda =2/p_0$, 
$\pm p_0$ being the polarization  at the two halves of the domain 
wall~\cite{Antal1999,Lancaster2010}. 
From Eq.~(\ref{cxx}) and Eq.~(\ref{I1}), these oscillations
also imply a current flow of magnitude $I_{s,n}^{(1)}= (J/\pi)\sin(\pi p_0)$.
In our case $p_0$=$1/2$.

\begin{figure}[htb]
\includegraphics[width=3.5in]{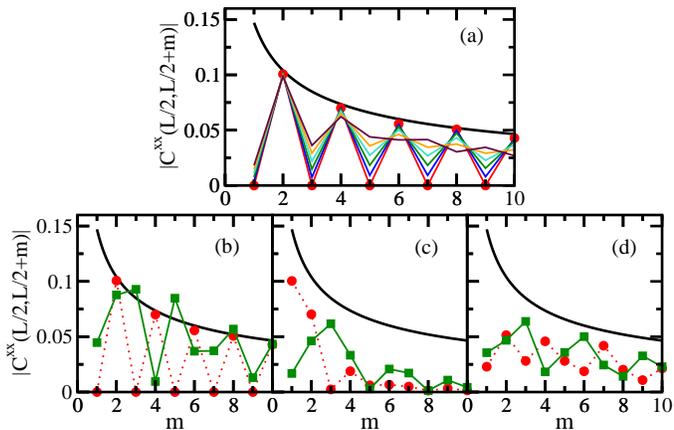}
\caption{(Color online.) Absolute value of the transverse spin correlation function, 
$|C^{xx}(L/2,L/2+m)|$ vs $m$ for time $t=10/J$ when only nearest-neighbor
couplings are present and for $t=6/J$ when next-nearest-neighbor
couplings are also present; $L=20$. For the three models, circle: $\Delta=0$ and square: $\Delta=0.3$.
Panels (a) and (b): integrable model with $\epsilon_{L/2+1}=0.0, \alpha=0.0$.
In panel (a) the results deviate monotonically from the case where $\Delta=0.0$
as the interaction increases: $\Delta =0.02, 0.04, 0.06, 0.08, 0.10$.
Panel (c): chaotic model with $\epsilon_{L/2+1}=0.5J, \alpha=0.0$.
Panel (d): chaotic model with $\epsilon_{L/2+1}=0.0, \alpha=0.4$.}
\label{fig1:Cxx}
\end{figure}

The equilibrium value of the $xx$-correlation function, $C_{eq}^{xx}(m)$,
is shown with a solid (black) line in all panels of Fig.~\ref{fig1:Cxx}.
The absolute
value of $C^{xx} (L/2,L/2+m)$ for the XX model
is depicted with circles in panels (a) and (b), and perfectly agrees 
with the analytical result.
For the XXZ model, as $\Delta$ increases, the results deviate gradually
from Eq.~(\ref{cxx_XX}), as illustrated in panel (a). For $\Delta>0.1$, it 
reaches a point where longer wavelengths become evident [panel (b)].
As will be shown in the
next sub-section, this increase in the wavelength can be associated with 
a decrease in the current as $\Delta$ is increased. 

The results for the chaotic systems are shown in panels (c) and (d).  
The overall decay of the 
correlation function is found to be more rapid  
than for the XX chain [Eq.~(\ref{cxx_XX})]. Moreover, both chaotic models show 
an overall increase in wavelength as $\Delta$ increases.   
As mentioned before, for the NN+NNN model, the simple 
correspondence between the nearest-neighbor correlation function 
($C^{xx}(n,n+1)$) and current no longer holds. Thus 
even though the wavelength of oscillations increases
for the NN+NNN model 
as $\Delta$ is increased from $0$ to $0.3$,
in this regime the current in fact also increases, as shown in Fig.~2 (c).

\subsection{Local spin current}

Spin and energy currents, $I_s$ and $I_E$ respectively, 
have been extensively studied in the context
of integrable periodic chains~\cite{Zotos1997,ZotosREVIEW1,ZotosREVIEW2,Heidrich2003}. 
$I_E$ commutes with the Hamiltonian of the XXZ system 
implying divergent thermal conductivity.
Although $I_s$ is conserved only for the XX model, it has been argued that
spin transport should also be anomalous in the XXZ case
when $S^z \neq0$~\cite{Zotos1997,Sirker2011}.

\begin{figure}[htb]
\includegraphics[width=3.5in]{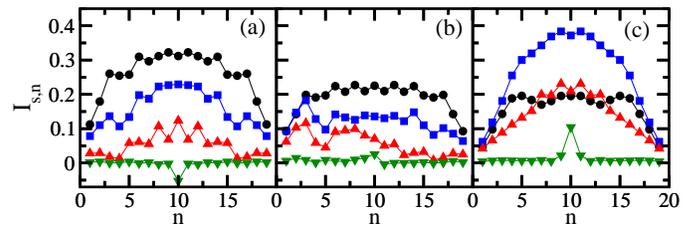}
\caption{(Color online.) Local current, 
$I_{s,n}$ [Eqs.~(\ref{I1}), (\ref{I2})] vs $n$ for time $t=10/J$ when only nearest-neighbor
couplings are present and for $t=6/J$ when next-nearest-neighbor
couplings are also present; $L=20$. 
Panels (a): integrable model with $\epsilon_{L/2+1}=0.0, \alpha=0.0$.
Panel (b): chaotic model with $\epsilon_{L/2+1}=0.5J, \alpha=0.0$.
Panel (c): chaotic model with $\epsilon_{L/2+1}=0.0, \alpha=0.4$.
Circle: $\Delta=0$; square: $\Delta=0.5$; 
up-triangle: $\Delta=1.0$; down-triangle: $\Delta=2.0$.}
\label{fig2:current}
\end{figure}

Numerical results for the local spin current, Eqs.~(\ref{I1}) and (\ref{I2}),
for the three models studied here are shown in Fig.~\ref{fig2:current}.
For NN couplings only, $I_s$ decreases 
as the  strength of the Ising interaction increases for both the 
integrable model [panel (a)] and the chaotic case with impurity [panel (b)].
For the integrable model, this behavior is expected.
Since the energy of the domain wall, $(L-3)\Delta/4$, 
increases with $\Delta$, the number of states capable of coupling to it decreases
and therefore the motion of the excitations becomes more restricted.
Equivalently, the eigenstates become spatially more localized.
For the system with impurity however, the addition of interaction
leads to the onset of chaos and corresponds to eigenstates that are spatially more delocalized.
(The relationship between the level of delocalization and the magnitude of the spin current
is further explored in Sec.~\ref{ipr}). At first sight, the existence of more
delocalized eigenstates seems to contradict the decrease of the current, but this decrease 
may be due to a crossover from ballistic to diffusive transport. 

The transport behavior of the three models considered here have been 
studied before. It is ballistic in the integrable domain, which occurs 
when $\alpha=0, \Delta\leq 1, \epsilon_n=0$ \cite{Zotos1997,Heidrich2003,Jung2007}
and also when 
$\alpha=0, \Delta=0, \epsilon_{L/2+1} \neq 0$ \cite{Barisic2009}, but 
it becomes diffusive for the chaotic models.
The relationship between non-ballistic transport and the onset of chaos
in the single impurity model with interaction
was investigated in Ref.~\cite{Barisic2009}. There it was shown that
spin and thermal conductivity relax within a time that scales with the
size of the chain. In the case of NNN couplings, the 
opening of a gap has been associated with diffusive 
transport~\cite{Alvarez200289,Heidrich2003,Heidrich2004,Langer2009},
but not much work exists that explicitly connects transport
behavior with the level of chaoticity of the system.

The model with NNN couplings and $\alpha=0.4$ is chaotic for any value of $\Delta \leq 1$.
The results for local current for this system, shown in panel (c),
are yet more surprising.  Here, the behavior of the current with $\Delta$ is non-monotonic.
It first increases
with $\Delta$, reaching a maximum level at $\Delta \sim 0.5$, and only then
starts diminishing with increasing $\Delta$. 
The level of spatial delocalization, on the other hand, decreases 
monotonically with $\Delta$  (see discussion in Sec.~IV)
and thus cannot justify the result for current.
A more thorough analysis of the level of chaoticity in the presence and
absence of Ising interaction, as well as how it relates to 
transport behavior, may shed some light on this scenario.

\subsection{Local magnetization}

\begin{figure}[htb]
\includegraphics[width=3.5in]{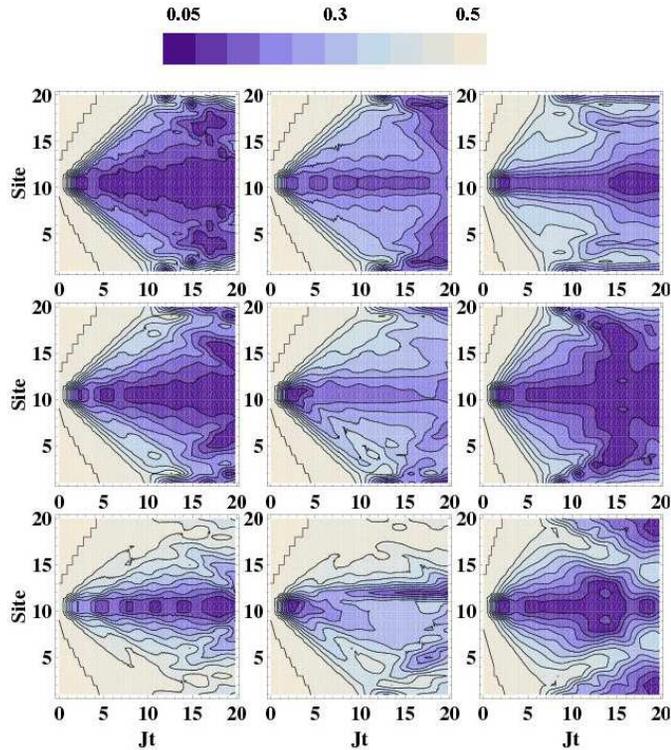}
\caption{(Color online.) 
Contour plots of the absolute value of the magnetization 
of each site $|M_n|$ [Eq.~(\ref{Mn})] vs time; $L=20$.
Left column: integrable model with $\epsilon_{L/2+1}=0.0, \alpha=0.0$.
Middle column: chaotic model with $\epsilon_{L/2+1}=0.5J, \alpha=0.0$.
Right column: chaotic model with $\epsilon_{L/2+1}=0.0, \alpha=0.4$.
Top panels $\Delta=0$; middle panels: $\Delta=0.5$; 
bottom panels: $\Delta=1.0$. The contour lines range from 
0.05 in the darkest region to 0.5 in the lightest region.}
\label{fig3:contour}
\end{figure}

\begin{figure}[htb]
\includegraphics[width=3.5in]{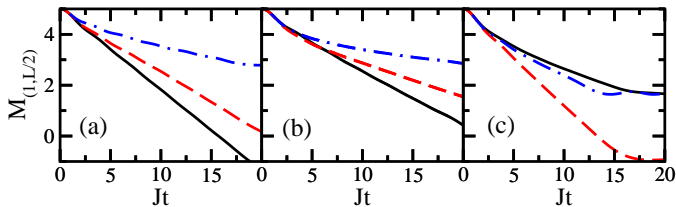}
\caption{(Color online.) 
Magnetization of the first half of the chain [Eq.~(\ref{M1})] vs time; $L=20$.
Panel (a): integrable model with $\epsilon_{L/2+1}=0.0, \alpha=0.0$.
Panel (b): chaotic model with $\epsilon_{L/2+1}=0.5J, \alpha=0.0$.
Panel (c): chaotic model with $\epsilon_{L/2+1}=0.0, \alpha=0.4$.
Solid line $\Delta=0$; dashed line: $\Delta=0.5$; 
dot-dashed line: $\Delta=1.0$.}
\label{fig4:magnetization}
\end{figure}

\begin{figure}[htb]
\includegraphics[width=3.5in]{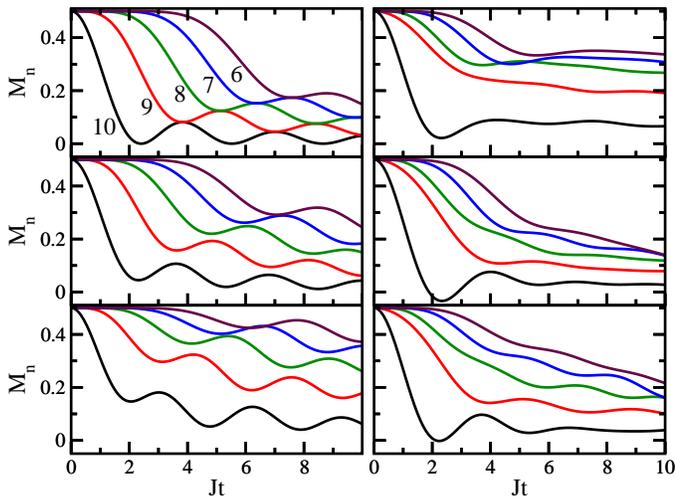}
\caption{(Color online.) 
Magnetization of site $n$ vs time; $L=20$.
Left panels: integrable clean model with $\epsilon_{L/2+1}=0.0, \alpha=0.0$.
Right panels: chaotic clean model with NNN couplings, $\epsilon_{L/2+1}=0.0, \alpha=0.4$.
In each panel, the curves range from site 10 to site 6, as indicated in the 
top left panel. Site 10 corresponds to the center of the domain wall;
the up-spin placed here is the first one to move.
Top panels $\Delta=0$; middle panels: $\Delta=0.5$; 
bottom panels: $\Delta=1.0$.}
\label{fig5:site_magnetization}
\end{figure}

The non-monotonic behavior of the  model with $\alpha \neq 0$ is reflected also in the
results for the local magnetization in Figs.~\ref{fig3:contour}, \ref{fig4:magnetization},
and \ref{fig5:site_magnetization}.

The contour plots in Fig.~\ref{fig3:contour} show the 
absolute value of the magnetization of each site vs time.
The interaction strength increases from top to bottom.
For the integrable model (left panels), as $\Delta$ 
increases, the motion of the excitations becomes more difficult
and it takes longer for the sites to reach very low absolute values of the magnetization.
A more complicated and asymmetric picture emerges in the
presence of an impurity (middle panels), but the trend is still the same.
However, when NN and NNN couplings are considered (right panels), 
as $\Delta $ increases from zero to 0.5,
the excitations spread much more 
efficiently and lead to an enhanced decay of the absolute values of local magnetization. 
As $\Delta$ is increased beyond $0.5$, the trend finally reverses and the
behavior becomes more similar to the other models, where the decay slows down
with the Ising interaction.

The local magnetization of the first half of the chain [Eq.~(\ref{M1})]
is shown in Fig.~\ref{fig4:magnetization} and reinforces the
observations from the contour plots.
The decay of $M_{(1,L/2)}$ is linear and the most rapid when $\Delta=0$
in the case of the integrable clean model [panel (a)] and the
model with a single impurity [panel (b)]. 
For these two cases, the decay slows down as the Ising interaction increases.
Panel (c), on the other hand, provides a clear
illustration of the  non-monotonic behavior of the NN+NNN model with $\Delta$.
Notice the abrupt and linear decay of $M_{(1,L/2)}$ when $\Delta=0.5$, 
which is to be contrasted with the slow and nonlinear decay for $\Delta=0$ and 1.

Another perspective to the results of the contour plots 
is provided in Fig.~\ref{fig5:site_magnetization},
which shows the behavior of the magnetization $M_n$ of site $n$ in time.
From this figure one verifies
the existence of two different velocities at play.
One is related to how fast the magnetization can reach very small values.
For the integrable model, $M_n$ reaches the lowest values faster when $\Delta$=$0$,
which justifies the steep slope of $M_{(1,L/2)}$ (solid line) 
in Fig.~\ref{fig4:magnetization} (a),
whereas for the chaotic model with NNN couplings, this speed is higher
for $\Delta$=$0.5$. 

The other velocity is related to the motion of the domain wall front. 
For the integrable case, this velocity increases
with $\Delta$. This is a well known result from Bethe Ansatz~\cite{KorepinBook}  
and bosonization~\cite{Sirker2006}, and has also 
been studied numerically in~\cite{Manmana2009}. A rough idea of the front
velocity may be obtained by looking at the 
first moment the magnetization reaches its minimum value. For the integrable model 
this happens earlier as the Ising interaction increases. Compare, for example, 
the time when site 10 reaches the lowest magnetization for $\Delta=0$ 
[top left panel] and  $\Delta =1$ [bottom left panel]: it is after
$2/J$ for the first and before $2/J$ for the latter. The same pattern is seen 
for all other sites. By passing an imaginary line through these minima, one sees that the slope
becomes steeper as $\Delta$ increases.
In the case of the chaotic model with NNN couplings, an evident
feature is the suppression of oscillations, but an estimate for the front
velocity becomes more difficult. Focusing on the magnetization of the
middle site 10, the velocity appears to also show a non-monotonic behavior.
Contrary to the integrable case, the front velocity 
appears to be slightly larger for $\Delta=0$ than for $\Delta=0.5$; it is only
for $\Delta > 0.5 $ that the behavior for the two models become
similar and the spin wave velocity increases again with interaction~\cite{KollathV}.

\section{Delocalization measure} \label{ipr}

In an attempt to understand the non-monotonic behavior of the model with NN and NNN couplings,
we study here the so-called inverse participation ratio (IPR).
For an eigenstate $|\psi_j\rangle$ written in the basis vectors $|\phi_k\rangle$ as 
$|\psi_j\rangle = \sum_{k=1}^{\text{dim}} c^k_j |\phi_k\rangle$, IPR is defined as

\begin{equation}
\mbox{IPR}_j \equiv \frac{1}{\sum_{k=1}^{\text{dim}} |c^k_j|^4} .
\label{IPR}
\end{equation}
This quantity measures the number of basis vectors that contribute to 
each eigenstate, that is, it measures
the level of delocalization of each eigenstate~\cite{Izrailev1990,ZelevinskyRep1996}.  
It is small when the state is localized and large otherwise.
Notice that the result depends on the choice of basis.
This choice is made according to the question
one is after. In studies of spatial localization, 
for example, the site-basis is the most appropriate one.
When the goal is to separate regular from chaotic behavior, the mean-field
basis~\cite{ZelevinskyRep1996}, corresponding to the eigenstates of the integrable 
limit, becomes more relevant.

We consider here the site-basis, since we are interested
in quantifying the spatial broadening of the domain wall.
For the three models and different values of the Ising interaction strength,
Table~\ref{table:ipr}
gives the value of IPR averaged over all eigenstates, $\langle \text{IPR}\rangle $; the 
number of states that participate in the evolution 
of the initial domain wall state, $|\Psi(t)\rangle = \sum_{j} C_j |\psi_j \rangle e^{-iE_jt}$,
with probability $|C_j|^2$ larger than $10^{-6}$; and the  IPR averaged over these
latter eigenstates, $\langle \text{IPR} \rangle_{|C_j|^2>10^{-6}} $. 
The purpose of these quantities is the following:
$\langle \text{IPR}\rangle $ quantifies the overall level of delocalization of the whole system
taking into account all $L!/[(L/2)!]^2$ eigenstates;
the number of states with $|C_j|^2>10^{-6}$ identifies the eigenstates that most
contribute to the evolution of the domain wall state; and finally
$\langle \text{IPR} \rangle_{|C_j|^2>10^{-6}} $ indicates whether these contributing
eigenstates have or have not the same level of delocalization of the whole system.

\begin{table}[ht]
\caption{Value of IPR averaged over all eigenstates;
number of states participating in the
evolution of the domain wall state, $|\Psi(t)\rangle = \sum_{j} C_j |\psi_j \rangle e^{-iE_jt}$, 
with probability larger than $10^{-6}$;
IPR averaged over these eigenstates that contribute with $|C_j|^2>10^{-6} $.
System size: $L=16$, subspace $S^z=0$.}
\begin{center}
\begin{tabular}{|c|c|c|c|c|}
\hline 
Model & & & & \\ 
($\alpha, \epsilon_{L/2+1}/J$) &  $\Delta $ & 
 $\langle \text{IPR} \rangle $ &  $|C_j|^2$'s $>10^{-6}$ &  
 $\langle \text{IPR} \rangle_{|C_j|^2>10^{-6}} $  \\
\hline 
\hline 
(0.0,0.0) & 0.0 & 3075.56 & 3054 &  3009.98 \\ 
\hline
(0.0,0.0) & 0.5 & 2358.02 & 2211 &  2333.44 \\
\hline
(0.0,0.0) & 1.0 & 1732.02 & 1139 &  1430.35 \\
\hline 
\hline  
\hline  
(0.0,0.5) & 0.0 & 1921.53 & 6696 &  1917.91 \\ 
\hline
(0.0,0.5) & 0.5 & 3679.49 & 6222 &  3737.60 \\
\hline
(0.0,0.5) & 1.0 & 3118.76 & 2868 & 2439.36 \\
\hline 
\hline 
\hline 
(0.4,0.0) & 0.0 & 3906.90 & 4378  & 4011.37 \\ 
\hline
(0.4,0.0) & 0.5 & 3684.89 & 2241 &  3624.25 \\
\hline
(0.4,0.0) & 1.0 & 2840.99 & 663 &  1670.35 \\
\hline
\end{tabular}
\end{center}
\label{table:ipr}
\end{table}

For the three models, the behavior is the same as $\Delta$ increases from 0.5 to 1.0:
the eigenstates become more localized. This is reflected in the dynamics
of the three systems: the excitations become
more localized, which decreases the magnitude of the spin current 
and slows down the decay of local magnetization (while 
the velocity of the domain wall front increases).
However, as $\Delta$ changes
from 0.0 to 0.5, this correspondence remains valid only for the integrable system, whereas
for the other two models orthogonal behaviors are seen:

(i) In the case of the system with impurity, the addition of interaction 
is followed by a crossover to chaos, which
justifies the substantial increase of the value of IPR.
The fact that the magnitude of the spin current decreases,
going contrary to delocalization, may be indicating a transition from
ballistic to diffusive transport, which should slow down the motion of the 
excitations.

(ii) The system with NN and NNN couplings is chaotic when $\Delta=0.0$ and also 
when $\Delta=0.5$. Based on the spatial contraction of the eigenstates that occurs when the
Ising interaction is turned on, one might expect a decline in the motion of the excitations. 
However, what is verified in Figs.~\ref{fig2:current}, \ref{fig3:contour}, \ref{fig4:magnetization},
and \ref{fig5:site_magnetization} is just the opposite. The interpretation of this counter-intuitive
behavior requires further studies.

It is interesting to compare the values of $\langle \text{IPR} \rangle $ and  
$\langle \text{IPR} \rangle_{|C_j|^2>10^{-6}} $. They are very similar for small
$\Delta$, but differ when $\Delta=1$. As mentioned before,
the energy of the domain wall increases with the strength of the Ising interaction.
At $\Delta=1$ the energy is already very large and close to the edge of the spectrum.
In systems with few-body interactions (ours have only two-body interactions), 
eigenstates with energy close to the edges of the spectrum are more 
localized~\cite{Santos2010PRE,RigolSantos}.
Thus, the eigenstates that contribute to the evolution of the domain wall state are  
fewer (the numbers in the column of $|C_j|^2>10^{-6} $ decrease significantly
from $\Delta=0.5$ to $\Delta=1.0$)
and also more localized 
than those participating in the evolution of generic initial states
($\langle \text{IPR} \rangle_{|C_j|^2>10^{-6}} <\langle \text{IPR} \rangle$).
This should affect the relaxation process of the 
domain wall state, which should become slower
for large $\Delta$.

\section{Conclusion} \label{concl}

We considered a one-dimensional spin-1/2 system in both integrable and chaotic regimes.
The integrable limit corresponded to a clean chain in the 
presence of nearest-neighbor couplings only. 
Chaos was induced (i) by adding a single-site
impurity field or (ii) by adding next-nearest neighbor couplings.
The time-evolution of an initial state in the form of a sharp domain wall
was studied.

The analysis of the dynamics focused on transverse spin correlation function,
local spin current, and local magnetization. 
In order to understand the behavior of these quantities,
we resorted
to delocalization measures; the motivation being that usually 
a good picture of the static properties of a system may
indicate what to expect for its time evolution. This approach served us well when 
dealing with the integrable model. Here, as the Ising interaction strength $\Delta $ 
increases, the level
of spatial delocalization of the eigenstates lowers and, as one might expect,
the magnitude of the spin current decreases and the 
decay of the absolute value of local magnetization slows down.

The same behavior was verified for the chaotic models for the case of large $\Delta$,
but for small Ising interaction, the level of delocalization changed in a way
contrary to the change in current and magnetization.
For model (i), the addition of interaction leads to the onset of chaos
and consequently the eigenstates further delocalize. The magnitude of the current,
however, decreased. We speculate that this is due to the transition from ballistic
to diffusive transport. For model (ii), the system is already chaotic at $\Delta=0$
and the eigenstates only shrink with the Ising interaction. 
Contrary to that, the motion of the excitations is enhanced, as attested
by the results for spin current and local magnetization.
Further studies are needed to clarify this behavior.

\begin{acknowledgments}
We thank Emil Prodan for stimulating discussions.
LFS thanks Vasile Gradinaru for introducing her to EXPOKIT.
AM was supported by NSF-DMR (Grant No. 1004589).
\end{acknowledgments}

\end{document}